\documentclass[useAMS,referee,usenatbib]{biom}

\usepackage{float}
\usepackage{pgf}
\usepackage{tikz}
\usetikzlibrary{positioning}
\usepackage{graphicx}
\usepackage{caption}
\usepackage{subcaption}
\usepackage{lscape}
\usepackage{multirow}
\usepackage[colorlinks,bookmarksopen,bookmarksnumbered,citecolor=blue,urlcolor=blue]{hyperref}
\usepackage{amsmath}

 \usepackage{colortbl}

\def\bSig\mathbf{\Sigma}

\newcommand{\VL}{{{\rm VL}}}
\newcommand{\Ee}{{{\rm E}}}

\newcommand{\dd}{{{\rm d}}}

\newcommand{\cum}{{\rm cum}}
\newcommand{\cumlag}{{\rm cumlag}}

\numberwithin{equation}{section}
\theoremstyle{plain}

\title[Dynamic models for estimating the effect of HAART on CD4 in observational studies]{Dynamic models for estimating the effect of HAART on CD4 in observational studies: application to the Aquitaine Cohort study and the Swiss HIV Cohort Study.}

\author{M. PRAGUE$^{1,2,3,4,*}$\email{mprague@hsph.harvard.edu}, D. COMMENGES$^{2,3,4}$, J.M. GRAN$^{5}$,
B. LEDERGERBER$^{6}$,\vspace{-0.25cm} \\
 \textbf{J. YOUNG$^{7}$, H. FURRER$^{8}$, R. THIEBAUT$^{2,3,4}$}\\
\vspace{-0.25cm}  $^1$ \textit{Harvard T.H. Chan School of Public Health, Boston, USA} \\
\vspace{-0.25cm} $^2$ \textit{Univ. Bordeaux, ISPED, F-33000 Bordeaux, France} \\
\vspace{-0.25cm} $^3$ \textit{INSERM, Centre INSERM U897-Epid\'emiologie et Biostatistique, F-33000, Bordeaux, France} \\
\vspace{-0.25cm} $^4$ \textit{INRIA (SISTM) Centre Recherche Bordeaux Sud-Ouest, Univ. Bordeaux, Talence, France} \\
\vspace{-0.25cm} $^5$ \textit{Oslo Center for Biostatistics and Epidemiology, Dept. of Biostatistics, Univ. of Oslo, Norway}\\
\vspace{-0.25cm} $^6$ \textit{Division of Infectious Diseases and Hospital Epidemiology, Univ. Hospital Zurich, Switzerland } \\
\vspace{-0.25cm} $^7$ \textit{Basel Institute for Clinical Epidemiology and Biostatistics, Univ. Hospital Basel, Switzerland} \\
\vspace{-0.25cm} $^8$ \textit{Department of Infectious Diseases, Bern Univ. Hospital and Univ. of Bern, Switzerland}   \\
}

\begin{document}

\date{{\it Received February} 2015. {\it Revised ???} ???.  {\it
Accepted ???} ???.}

\pagerange{\pageref{firstpage}--\pageref{lastpage}}
\volume{64}
\pubyear{2015}
\artmonth{December}

\doi{10.1111/j.1541-0420.2005.00454.x}

\label{firstpage}

\begin{abstract}
Highly active antiretroviral therapy (HAART) has proved efficient in increasing CD4 counts in many randomized clinical trials. Because randomized trials have some limitations (e.g., short duration, highly selected subjects), it is interesting to assess it using observational studies. This is challenging because treatment is started preferentially in subjects with severe conditions, in particular in subjects with low CD4 counts. This general problem had been treated using Marginal Structural Models (MSM) relying on the counterfactual formulation. Another approach to causality is based on dynamical models. First, we present three discrete-time dynamic models based on linear increments (LIM): the simplest model is described by one difference equation for CD4 counts; the second has an equilibrium point; the third model is based on a system of two difference equations which allows jointly modeling CD4 counts and viral load. Then we consider continuous time models based on ordinary differential equations with random effects (ODE-NLME). These mechanistic models allow incorporating biological knowledge when available, which leads to increased power for detecting treatment effect. Inference in ODE-NLME models, however, is challenging from a numerical point of view, and requires specific methods and softwares. LIMs are a valuable intermediary option in terms of consistency, precision and complexity. The different approaches are compared in simulation and applied to HIV cohorts (the ANRS CO3 Aquitaine Cohort and the Swiss HIV Cohort Study).\\
\end{abstract}


\maketitle

\section{Introduction}
\label{sec1}

Assessing the effect of a treatment in observational studies is useful because randomized clinical trials  often have short durations and include highly selected subjects.
 This is challenging because the treatment may change, and covariates history of a subject up to time $t$ may influence treatment given after $t$, and may also influence the outcome of interest, which induces a time-dependent confounding. 
 For instance, one may wish to assess the effect of antiretroviral therapy in HIV infected subjects. As CD4+ T-lymphocytes (CD4, in short) are the main target cells of the HIV virus, it is possible to assess the effect of a treatment on the blood concentration of these cells: CD4 counts are measurements of this concentration. In observational studies, however, the decision to start an antiretroviral therapy may depend on CD4 counts as well as on other covariates. In this setting, it has been demonstrated that a conventional regression analysis leads to biased estimates of the treatment effect, typically underestimating it, and possibly (wrongly) indicating a negative effect. This  is called ``confounding by indication'' \citep{walker1996confounding}.

 The marginal structural models (MSM) \citep{Robins2000} have been proposed for treating this problem; this is based on choosing a causal model in terms of potential responses (which are often counterfactual) to the different treatment histories. The parameters of a MSM can be estimated through a weighted approach but other methods exist \citep{petersen2006assessing}. The weights are the inverse probability (IP) of treatment attribution and are obtained through a ``treatment model'' which includes the covariates linked to the outcome. Because data are correlated, we use an IP weighted generalized estimating equation (GEE). This approach has been applied by \cite{Hernan2002} and by \cite{Cole2005}  for estimating the effect of zidovudine and of highly active antiretroviral therapy (HAART) on CD4 count. \cite{cole2007determining}, \cite{sterne2005long} and  \cite{Cole2010} used it for estimating the effect of HAART on viral load and on AIDS or death.

An alternative approach which does not use the potential responses representation is to use dynamic models.
 The dynamic approach to causality has been pioneered by \cite{Granger1969}, \cite{Aalen1987}, and further developed by \cite{Didelez2008a}, \cite{Commenges2009}, \cite{Gegout-Petit2010} and \cite{Eichler2010}.  Assumptions needed for a causal interpretation of dynamic models have been presented in \cite{Arjas2004} and \cite{Commenges2015}. Dynamical models in discrete time, and in particular Linear Increment models (LIM), have been proposed by \cite{diggle2007analysis} and \cite{HoffFLIM2014}; \cite{Aalen2012a} have suggested that such models can be useful for studying HAART effect on CD4 counts or viral load.
Discrete-time models, however, may not be completely satisfactory because the processes of interest most often live in continuous time.
Systems of differential equations in continuous time can also be used to model the interaction between HIV and CD4 cells populations. Models based on differential equations, called ``mechanistic'', considerably helped in understanding some important features of the infection: see \cite{perelson2002modelling} for a review. In our setting, it is possible to model the treatment effect from a biological perspective. Introducing random effects allows analyzing a sample of subjects with different parameter values without too much increasing the number of parameters \citep{wu2005statistical,guedj2007maximum,lavielle2011maximum}.
Up to now, mechanistic models have been used to analyze data from clinical trials.
Using mechanistic models to estimate the effect of HAART based on data of large observational cohorts is a possibility, that to our knowledge has never been attempted.

The aim of this paper is to propose dynamic models in discrete  and continuous time for assessing the causal effect of a treatment on a marker in observational studies. Specifically, we aim at estimating HAART effect in HIV infected patients. We present several possible dynamic models, as well as MSM models, and compare them using simulations and real data. In Section 2, we present the statistical models: the naive model, the MSM models, the discrete-time dynamic models and a mechanistic model. In Section 3, we compare the results of these models in simulation, where the data are generated from a complex mechanistic model. Section 4 is the application on the data of two cohorts of HIV infected patients: the Swiss HIV Cohort Study (SHCS) and the ANRS CO3 Aquitaine cohort. Section 5 concludes.

\vspace{-0.5cm}
\section{Modeling the treatment effect in observational studies}
 \label{part:formalization}
\subsection{Notations and the naive model}

We denote the value of a physiological marker $Y$ for subject $i$ at time $t$ by $Y^i_t$. In this section we omit the superscript $i$ for simplicity.
The value of a treatment given at time $t$ is denoted by $A_t$. For sake of simplicity, we only model two treatment states: $A_t=0$ when treatment is not given, and $A_t=1$ when treatment is given at time $t$, and we assume that once initiated, the treatment is not interrupted; generalization to different treatment levels is possible. If treatment is started at time $t$ then $A_t=1$ and $A_{t-1}=0$.
In our application $Y_t$ is the CD4 counts, $A_t$ is treatment (HAART) attribution which is binary.
We use overbars to represent histories of the processes: for instance $\bar A_t=(A_0, A_1, \ldots, A_t)$. We denote by $\cum (\bar A_{t})$ the cumulative time under treatment until time $t$. Since A is binary we can write $\cum (\bar A_{t})=\sum_{k=1}^t A_k$

In the absence of confounding by indication, a regression of $Y_t$ on the history of treatment would give the effect of treatment on the marker. The simplest model would be to regress $Y_t$ on $\cum (\bar A_{t-1})$.  It has been noted however that a piecewise linear regression model allowing a change of treatment effect after one year was better suited \citep{Cole2005}. Thus, the naive model that we consider is our \textbf{Model 1}:
\begin{eqnarray}
\Ee (Y_t |  \bar A_{t-1} )= \beta_0 + \beta_1 \cum (\bar A_{t-1})+ \beta_2 \cumlag (\bar A_{t-1}). \label{eq:Model1}
\end{eqnarray}
where $\cumlag (\bar A_{t})$ is the cumulative time under treatment up to time $t$ minus one year: $\cumlag (\bar A_{t})=\max(0,\cum (\bar A_{t})-1)$ (with the convention $\bar A_{t}=0$ for $t<0$).
The $\beta$'s are estimated by conventional GEE \citep{liang1986longitudinal} because it is very likely that the $Y_t$'s are positively correlated, and because we are interested in the population average \citep{hubbard2010gee}. A working correlation structure has to be chosen; in presence of time-varying covariates, the independence model should be chosen, otherwise results could be biased, as shown by \cite{pepe1994cautionary}. Thus an identity working correlation matrix is used for all GEE models.

\vspace{-0.5cm}
\subsection{Marginal structural models (MSM)}

Since treatment is given to subjects with low CD4 counts,  treated subjects tend to have low CD4 counts. Thus, the true value of parameter $\beta_1$ in {Model 1} cannot be interpreted as the causal effect. The MSM have been designed to estimate the causal effect of a treatment in such a case.
It is assumed that to each particular value $\bar a_t$ of $\bar A_t$, a potential outcome $\bar Y_t(\bar a_t)$ is associated; $\bar a_t$ can be called ``treatment history'' or ``treatment trajectory''. This means that if a subject had (possibly contrary to the fact) treatment trajectory $\bar a_t$, his outcome would be $\bar Y_t(\bar a_t)$. A model is postulated to describe how the potential outcomes vary as a function of the different treatment trajectories. Then it has been shown that the parameters of this model, called ``causal parameters'', can be estimated with a suitably weighted GEE. The weights are inverse-probability-of-treatment (IPT). The probability of treatment at time $t$ depends on the history up to time $t$ of a vector of variables $L$; this history is denoted $\bar L_t= (L_0 \in \bar L_t)$; $L_t$ may include $Y_t$ itself, as well as other variables such as viral load or other biological markers linked to the infection. The probability of treatment is estimated using a treatment model (generally a logistic model) for each time and the weights are the product over time of these probabilities; one often use stabilized weights as in Equation (\ref{stabweights}.
The causal parameters can be estimated consistently if all the confounders (factors influencing both the outcome of interest and treatment attribution) have been taken into account.
Extension to the case with censoring has also been developed \citep{Cole2005,cole2008constructing}. However, the most important correction is generally for the probability of treatment \citep{ko2003estimating}.
\cite{Hernan2002} and \cite{Cole2005} proposed benchmark models also adjusting for confounders such as time ($t$) and baseline value of the biomarker ($Y_0$) in the regression. The model for potential outcomes that we consider is the same for our \textbf{Model 2} and \textbf{Model 3} (which will differ by the treatment model); it is:
\begin{eqnarray}
\Ee (Y_t(\bar a_t) | \bar a_{t-1},Y_0)=  \beta_0 + \beta_1 \cum (\bar a_{t-1})+ \beta_2 \cumlag (\bar a_{t-1}) +\beta_3 t +\beta_4 Y_0. \label{eq:Model23}
\end{eqnarray}

For estimating the parameters of this model, we used GEE with stabilized IPT weights defined as:
\begin{eqnarray}
SW(t)= \prod_{k=1}^{t}\frac{Pr(A_{k}=1| \bar A_{k-1}, L_0)}{Pr(A_k=1| \bar A_{k-1}, \bar L_k)}, \label{stabweights}.
\end{eqnarray}
where the probabilities at time $k$ are estimated for every subject from  logistic regressions depending on $\bar L_k$ ($L_0 \in \bar L_k$). We tried two different treatment models. We defined the subsets $L$ for treatment model in \textbf{Model 2} as baseline and time-varying CD4 count in class ($< 200$, $[200;400]$, $>400$) only. We extended this list for \textbf{Models 3} to the main potential confounders: viral load in categories ($<401$, $401-10000$ and $>10000$), and an indicator of  undetectable viral load.
In Models 1, 2 and 3, the effect of treatment on  CD4 counts during the first year is given by $\beta_1$ and the effect after one year of treatment is given by $\beta_1+\beta_2$.

\vspace{-0.5cm}
\subsection{Discrete-time dynamical linear increment models (LIM) }\label{sec:dynmod}
When using discrete-time dynamical models, we regress the change in the marker of interest, that is, $Z^i_t=Y^i_t-Y^i_{t-1}$.  This fits well with a causal thinking which considers that the change of a process depends on its present, and possibly past, state. The independence assumption for the $Z^i_t$'s is much more acceptable than for the $Y^i_t$'s. However, there remains an inter-subject variability that we  model by a random effect $b_i$, assumed normally distributed with zero expectation. Thus, these models specify the distribution of $Z^i$ conditional on the $b_i$. We propose three of these models; the simplest is \textbf{Model 4}:
\begin{eqnarray}
Z^i_t=  \beta_0 + \beta_1 A^i_{t-1}+ \beta_2 A^i_{t-2}+b_i+\varepsilon^i_t, \label{eq:Model4}
\end{eqnarray}
where the $\varepsilon^i_t$'s are i.i.d. normally distributed variables with zero expectation, and the $b_i$'s are normal random effects. This model can be easily fitted since this is a linear mixed-effects model.
In order to account for non equally spaced measurement of biomarkers, we extend the notation to obtain a model which has approximately the same meaning: it is natural to think that the change is proportional to the time elapsed between two observations, that we note $\Delta^i_t=c^i_t-c^i_{t-1}$, where $c^i_t$ is the $t^{th}$  calendar time of observation since baseline measure of subject $i$. This extended model is obtained by redefining the increment as  $Z^i_t=\frac{Y^i_t-Y^i_{t-1}}{\Delta^i_t}$.
In Models 4, the effect of treatment on the CD4 counts during the first year is approximated by $\beta_1 \bar \Delta_t$, and the effect after one year of treatment  by $(\beta_1+\beta_2) \bar \Delta_t$, where $\bar \Delta_t$ is the mean of all the $\Delta^i_t$.

 Many deterministic dynamical models have equilibrium points; similarly many stochastic dynamical models tend toward a stationary process: this property fits very well with the behavior of biological systems since concentrations of many molecules or cells have a tendency to return around the same value, a property called ``homeostasis''. The difference equation of the type $Y_t-Y_{t-1}=\gamma_0+\gamma_1 Y_{t-1} + \varepsilon_t$ corresponds to an autoregressive model of order one, noted  AR(1): $Y_t=\gamma_0+ \gamma' Y_{t-1} + \varepsilon_t$ with $\gamma'=(\gamma_1+1)$. It is well known that if $|\gamma'|<1$ this process converges toward a stationary process (in discrete time) with expectation $\Ee(Y_{t})=-\frac{\gamma_0}{1-\gamma'}=-\frac{\gamma_0}{\gamma_1}$; this is always defined unless $\gamma_1=0$, as is the case in Model 4 which does not have a finite stationary expectation. The condition amounts to $-2<\gamma_1<0$, and to get a positive stationary expectation we must have $\beta_0>0$. When using a model which has this convergence property, it may not be necessary to have a two-slope model, so we define \textbf{Model 5} as:
\begin{eqnarray}
 Z^i_t=  \beta_0 + \beta_1 A^i_{t-1}+ \beta_2 Y^i_{t-1} + b_i+\varepsilon^i_t. \label{eq:Model5}
\end{eqnarray}
If $-2<\beta_2<0$ and $\beta_0 + \beta_1>0$, $Y$ tends to a stationary process with expectation $-\frac{\beta_0 + \beta_1}{\beta_2}$ for treated patients.

A more realistic modeling of CD4 counts is to take viral load into account. Here, we make a step toward mechanistic models because we know that the virus concentration and the CD4 concentration are two inter-related processes. Thus, we propose \textbf{Model 6} based on a system of two difference equations:
\begin{eqnarray}
\small
\left \{ \begin{array} {lcl}
Z^{i}_t &=&\beta_0 + \beta_1 A^i_{t-1} + \beta_2 Y^i_{t-1} + \beta_3 \VL^i_{t-1} +b_i+\varepsilon^1_{it},\\
W^{i}_t&=&\alpha_0+ \alpha_1 A^i_{t-1}  +\alpha_2 Y^i_{t-1}+ \alpha_3 \VL^i_{t-1}+d_i+\varepsilon^2_{it}.
\end{array} \right. \label{eq:Model6}
\end{eqnarray}
where  $W^i_t=\frac{\VL^i_t-\VL^i_{t-1}}{\Delta^i_t}$, with $\VL^i_t$ the viral load at time $t$, $d_i$ and $b_i$ are normally distributed independent random effects, and $\varepsilon^1_{it}$ and $\varepsilon^2_{it}$ are normal i.i.d. error variables with zero expectation.
One year and subsequent years increase of CD4, as well as the long term change, are easily computed by solving the difference equations numerically.
 For testing whether the treatment has an effect, it is convenient to test the hypotheses $\beta_1=0$ and $\alpha_1=0$, by Wald tests for instance.
As a reviewer noted, an interesting question is the correspondence between these LIM models and MSMs.  Appendix shows that under some (rather strong) assumptions, Model 4 allows estimating the same causal parameters as Models 2-3.

\vspace{-0.5cm}
\subsection{Continuous Dynamical models, Mechanistic Models (ODE-NLME)}

In reality, biomarkers processes live in continuous time. 
We  use the ``target cells model'' that proved to provide a good fit and to have good prediction abilities \citep{prague2013dynamical}.
The combination of the target cells model, a model for inter-individual variability of the parameters, and an observation model specifies our \textbf{Model 7}.

\begin{figure}[ht!]
        \centering
        \begin{subfigure}[b]{0.4\textwidth}
                \includegraphics[width=\textwidth]{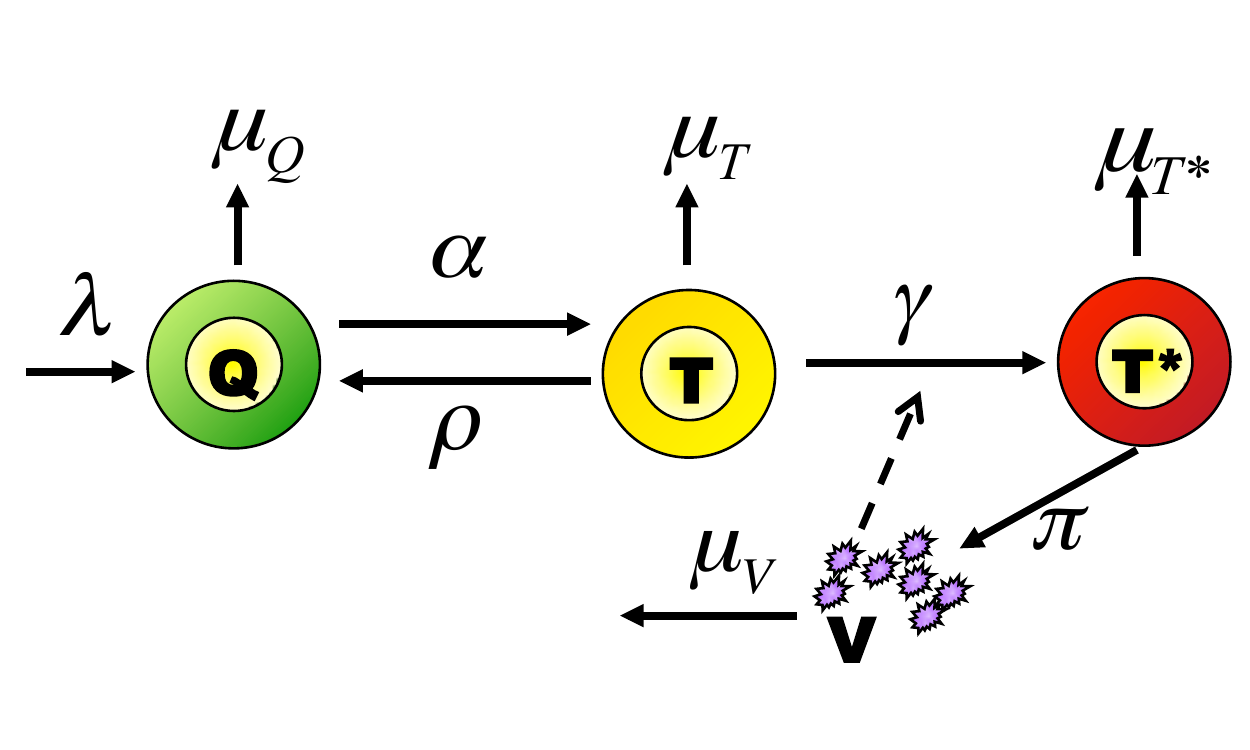}
                \caption{Target cell model }
                \label{fig:mecaPrague}
        \end{subfigure}%
        $\qquad$~ 
        \begin{subfigure}[b]{0.6\textwidth}
                \includegraphics[width=\textwidth]{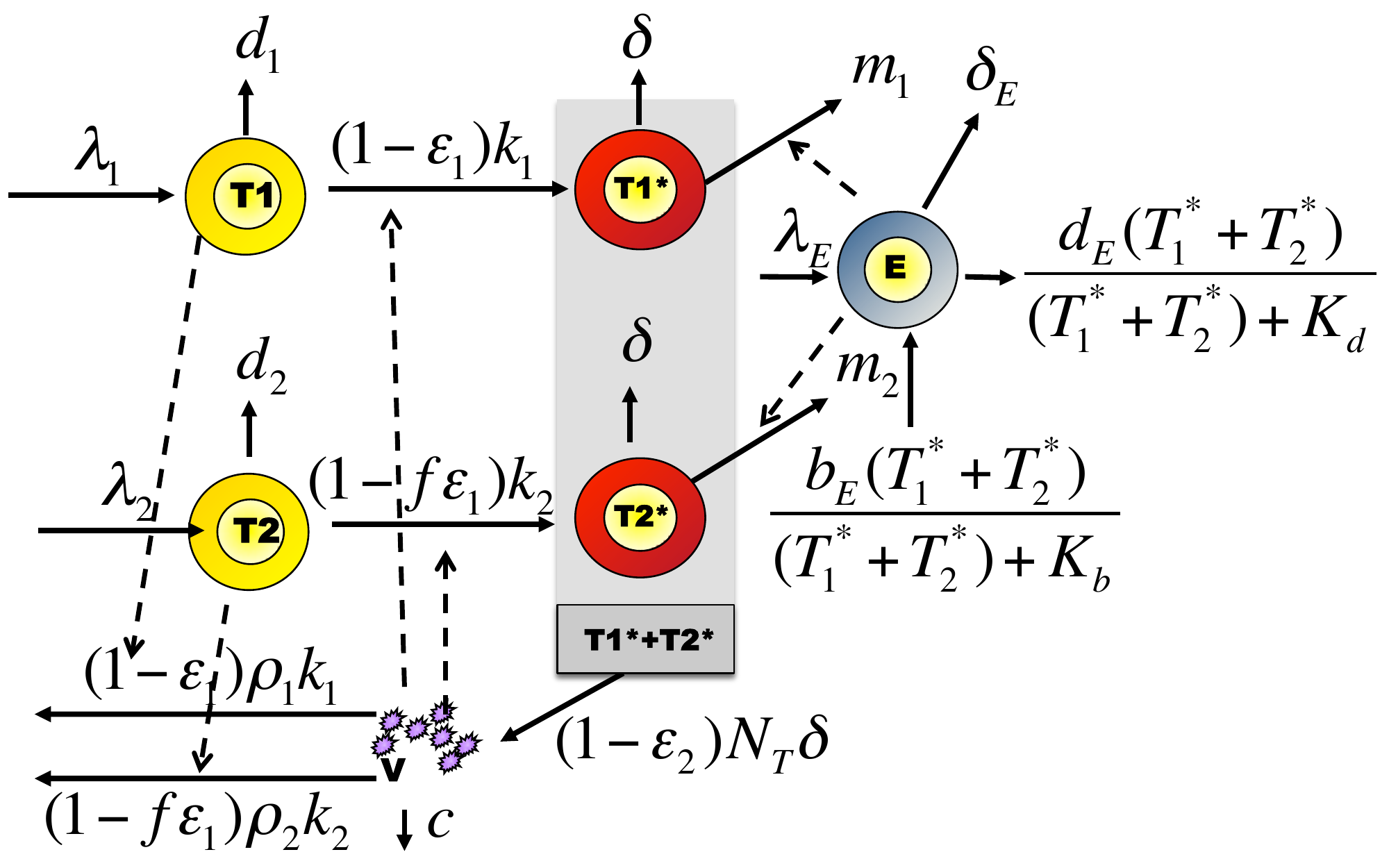}
                \caption{\cite{adams2005hiv} model.}
                \label{fig:mecaAdams}
        \end{subfigure}
        \caption{Mechanistic models for HIV dynamics. Type of cells of interest are viruses ($V$), effector cells $E$ and CD4 cells which may be quiescent ($Q$), target cells ($T$, $T_1$, $T_2$) or infected  ($T^*$, $T^*_1$, $T^*_2$). Parameters are defined in Table \ref{tab:mecaAdams}. }\label{fig:meca}
\end{figure}

{\bf Biological system.}
We know that only infected cells ($T^*$) can produce viruses ($V$).
The target cells model distinguishes between uninfected quiescent cells ($Q$) and target cells ($T$). The instantaneous change of concentrations of these populations at time $t$, for all real value of $t>0$, is given by the ODE system:
\begin{equation}
\small
\left \lbrace\begin{array} {lcl}
\frac{\dd Q^i_t} {\dd t}&=& \lambda^i + \rho^i T^i_t - \alpha^i Q^i_t- \mu^i_{Q^i} Q^i_t,\\
\frac{\dd T^i_t} {\dd t} &=& \alpha^i Q^i_t  -\gamma^i T^i_tV^i_t - \rho^i T^i_t - \mu^i_{T}T^i_t, \\
\frac{\dd {T^{*}}^i_t)} {\dd t} &= & \gamma^i  T^i_tV^i_t - \mu^i _{T^{*}} {T^{*}}^i_t, \\
\frac{\dd V^i_t} {\dd t} & =& \pi^i {T^{*}}^i_t - \mu^i_{V}V^i_t. \\
\end{array} \right.  \label{eq:Model7}
\end{equation}
The system is graphically represented in Figure \ref{fig:mecaPrague}.
Here, the parameters have biological meanings: $\lambda$ is the production rate of new CD4 cells, the $\mu$'s are death rates, $\alpha$ and $\rho$ are transition rates between quiescent and target cells, $\pi$ is the rate of production of virions by infected cells, and $\gamma$ is the infectivity parameter. The model assumes that the rate of infection of target cells is $\gamma V_t$.

\begin{table}[ht!]
\caption{Meaning of parameters in the dynamical models presented in Figure \ref{fig:meca}. The upper part gives prior means and standard deviations for normal a priori distributions used for estimation of mechanistic parameters in Model 7 for the ``Target cell model''. The lower part gives parameter values used for data simulation from the \citet{adams2005hiv} model.}
\begin{center}
\small
\begin{tabular}{lllcccc}
  \hline
  &&&\multicolumn{4}{c}{Normal priors used for analysis$^*$}\\
    &&&\multicolumn{4}{c}{on the log value of the parameter}\\ \cline{4-7}
Name & Description &&\multicolumn{2}{c}{mean}&\multicolumn{2}{c}{sd. } \\
  \hline
   $\lambda         $&    Natural production rate &$\frac{cells}{\mu L.day}$&   \multicolumn{2}{c}{  2.55}&        \multicolumn{2}{c}{    1.90 }\\
$\mu_{T^*}       $&    Natural death rate of $T^*$ cells & $\frac{1}{day}$&  \multicolumn{2}{c}{  -0.05}&    \multicolumn{2}{c}{        0.68}\\
$\mu_Q           $&     Natural death rate of $Q$ cells & $\frac{1}{day}$&   \multicolumn{2}{c}{   -9.00}&    \multicolumn{2}{c}{        1.00}\\
$\alpha          $& Transition rates between Q and T cells    & $\frac{1}{day}$ &   \multicolumn{2}{c}{        -4.00}&        \multicolumn{2}{c}{    2.00}\\
$\rho            $&   Transition rates between T and Q cells     & $\frac{1}{day}$& \multicolumn{2}{c}{     -4.34}&         \multicolumn{2}{c}{   1.38}\\
$\mu_T           $&       Virions natural death rate & $\frac{1}{day}$&      \multicolumn{2}{c}{       -2.59}&      \multicolumn{2}{c}{      0.34} \\
$\gamma          $&   Infectivity parameter   & $\frac{\mu L}{day}$ &      \multicolumn{2}{c}{        -5.76}&   \multicolumn{2}{c}{         4.02  } \\
$\pi             $&    Rate of production of virions by infected cells    & $\frac{1}{day}$&   \multicolumn{2}{c}{        4.04}&         \multicolumn{2}{c}{   2.66 } \\
$\mu_V           $&        Natural death rate of $T^*$ cells & $\frac{1}{day}$ &      \multicolumn{2}{c}{        2.83}&      \multicolumn{2}{c}{      0.68 }    \\
  \hline
  &&&\multicolumn{4}{c}{Parameter Value~used for simulations }\\
    &&&\multicolumn{4}{c}{for each population (X)\textdagger }\\ \cline{4-7}
Name & Description &Units& Type 1 & Type 2 & Effectors & Virus \\
  \hline
$\lambda_X$& Natural production rate &$\frac{cells}{mL.day}$& $5000$& $31.98$&$1.0$& -\\
(1-$\epsilon_X)$& Treatment efficacy & no unit & 50\% &83\% & - &- \\
$d_X$& Natural death rate & $\frac{1}{day}$ & 0.01&0.01 & 0.25  &-\\
$\delta_X$& Infection-induced death rate & $\frac{1}{day}$ & 0.7 &0.7 & 0.1  &-\\
$\rho_X$& Number of virions infecting a cell  & $\frac{virions}{cells}$ & 1&1 & - &-\\
$m_X$& Immune-induced clearance rate & $\frac{mL}{cells.day}$ & $1 \times 10^{-5}$&$1 \times 10^{-5}$ & - &-\\
$k_X$& Infection rate & $\frac{mL}{virions.day}$ & $8 \times 10^{-7}$&$1 \times 10^{-4}$ & -&- \\
$c$ & Virions natural death rate & $\frac{1}{day}$  & - & - & - & 13 \\
$N_T$ & Virions production per infected cells & $\frac{virions}{cells}$  & - & - & - & 100 \\
$K_b$ & Saturation constant cells birth & $\frac{cells}{mL}$  & - & - & 100& - \\
$K_d$ & Saturation constant cells death & $\frac{cells}{mL}$  & - & - & 500 & - \\
$b_E$ & Infection-induced birth rate for E cells & $\frac{1}{day}$  & - & - & 0.3 & - \\
   \hline
       \multicolumn{7}{l}{$^*$ Reference and explanation for these choices can be found in \cite{prague2012treatment}. }\\
    \multicolumn{7}{l}{ \textdagger For each simulated patient, every parameter got a random effect leading to 50\% coefficient of variation} \\
    \hline
\end{tabular}
\end{center}
\label{tab:mecaAdams}
\end{table}

 {\bf Inter-individual variability.}
 The model for inter-individual variability of the parameters is a mixed-effect model for the log-transformed parameters denoted with a tilde.
 In this application, based on \cite{prague2012treatment}, two random effects $u^i_{\lambda}$ and $u^i_{\mu_{T^*}}$ are introduced:  $\tilde \lambda^i=\tilde \lambda_0+u^i_{\lambda}$ and $\tilde \mu^i_{T^*}=\tilde \mu_{T^*0}+u^i_{\mu_{T^*}}$.
Biologically, the causal effect of treatment
can be modeled as an effect on the infectivity parameter $\gamma$. The parameter $\gamma$ consequently depends on $t$ through $A_t$:
\begin{equation}
\label{modeldose}\tilde \gamma^i(t)= \tilde \gamma_0+\beta A^i_t,
\end{equation}
where we expect $\beta<0$, so that the treatment decreases the infectivity of the virus.

{\bf Observation model.}
One important consequence of using continuous time models is that we must distinguish between the biological system which lives in continuous time and observations which are made at discrete time.
To make an additive model for measurement error acceptable, we use 4th-root transformation for CD4 and a log$_{10}$ transformation for the viral load respectively. Thus, the observation model is:
\begin{equation}\label{observationmodel} (Y^i_j)^{1/4}=[Q^i_{t_{ij}}+T^i_{t_{ij}}+{T^*}^i_{t_{ij}}]^{1/4}+\varepsilon^1_{ij} ~~{\mbox ; } ~~ \log_{10}\VL_{ij}= \log_{10} V^i_{t_{ij}}+\varepsilon^2_{ij},\end{equation}
where $\varepsilon^1_{ij}$ and $\varepsilon^2_{ij}$ are measurement errors,  independently normally distributed.

{\bf Inference.} Inference is much more complex and computationally demanding than in discrete-time models; it is based on a penalized maximum likelihood approach; in order to avoid identifiability issues, we used {\em a priori} knowledge on mechanistic parameters: the priors are set  according to past estimates of the biological parameters in the literature \citep{prague2012treatment}, and are given in Table \ref{tab:mecaAdams}. A special Newton-like algorithm has been implemented in the NIMROD program \citep{Prague2013NIMROD}.
Assessing the long-term treatment effect in Model 7 is possible by analytically computing the equilibrium point. One year and subsequent years increase of CD4 after treatment initiation can be computed by solving the ODE system for given values of the random effects. The marginal effect can be computed as the mean of the individual effects in the population. The infectivity parameter gives an indicator of the effect of treatment, and a Wald test can be used to test the no-effect hypothesis ``$\beta=0$'' .


\section{Simulation study}
\label{part:simul}

 We simulated composite data with the \cite{adams2005hiv} model, which is much more complex than Model 7 (see Web-Supplementary Material A4): it includes two populations of target cells and a population of immune effectors (such as cytotoxic T-lymphocytes); see Figure \ref{fig:mecaAdams} and Table \ref{tab:mecaAdams}. Parameters have inter-individual variability modeled by drawing them from a normal law (with mean values listed in Table \ref{tab:mecaAdams} and variances chosen to obtain a variation coefficient of 50\%).
By controlling the value of random effects, we ensure that the steady state baseline distributions of CD4 counts and viral load are consistent with the baseline values distributions found in Aquitaine cohort and SHCS dataset. See Appendix A1 in Web-supplementary Material for details. We generated observations every 3 months; the standard deviations of the measurement errors are $\sigma_{VL}=0.6$ and $\sigma_{CD4}=0.1$. Viral load was artificially made undetectable at the level of 50 copies/mL. Treatment assignment was done by simulating a CD4 count assessment at every visit (every 3 months) and by fixing a probability of treatment attribution depending on the observed CD4 count. We took empirical probabilities from the Aquitaine cohort and SHCS dataset: treatment was attributed in 2\%, 28\% or 47\% of the cases if CD4 count was $>400$, $[400,200]$ or $<200$. Neither confounder nor drop-out was considered. We simulated n=200 and n=1500 patients.
Table \ref{tab:descriptionDataset} gives a general description of both simulated and real data sets: no major inconsistency in descriptive statistics values appears between simulated and real cohort data sets. We define the ``average causal effect in treated patients'' as the mean difference between the observed CD4 according to the observed treatment initiation and the counterfactual CD4 under no treatment initiation. The result of this computation was a 350 cells increase of CD4 after 1 year, a 362 cells increase after 2 years and an overall increase of 370 CD4 cells after an infinite (large) time. Technical implementation details and code for analysis are described in Web-Supplementary Material C.



 Table \ref{tab:resultsimul} presents the estimates for Models 1-7 on the simulated datasets.
 (see detail in Web-Supplementary Material A2 and A3).
 All results and conclusions are similar in small and large samples. The naive Model 1 largely underestimated the treatment effect.
This was corrected by the MSM Models 2 and 3.
Model 4 also yielded good estimates of the mean causal effect in treated patients both for the first year and subsequent years.
However, we can notice that long-term increase of CD4 is infinite in Models 1 to 4.
Models 5-7 exhibit an equilibrium point which makes it possible to consider the long-term causal effect of treatment. All dynamic models gave a correct estimate of the long-term effect of the treatment.
The initial increase in CD4 during the first year was not correctly caught by Model 5. Models 6 and 7 which both incorporate the dynamics of viral load gave a correct estimation of the increase of CD4 count.
Even if all models found a significant effect of treatment on CD4 counts in the first year, the Z-statistics for the no-effect hypothesis are larger for dynamical models than for the GEE-based models, indicating more power to reject the null hypothesis. Sandwich estimators were used for the standard errors (SE) for GEE method; adjusting for uncertainty in the estimation of weights would lead to larger SE and thus would not impact this conclusion. 

\begin{table}[ht!]
\caption{ Estimated treatment effect on CD4 counts from simulated data: Model 1: Naive regression; Model 2:  MSM with simple treatment model; Model 3:  MSM with more complete treatment model; Model 4: simple dynamic model; 5: autoregressive model; Model 6: bivarariate dynamic model; Model 7: mechanistic model.}
\begin{center}
\begin{tabular}{llccccccc}
  \hline
   & &\multicolumn{7}{c}{Simulated Dataset with \cite{adams2005hiv} model} \\
  & & \multicolumn{3}{c}{n=200}&$\quad$&\multicolumn{3}{c}{n=1500} \\\cline{3-5}\cline{7-9}
  Model & $\beta$ treatment$^*$&   Effect &  Sd. & Z-stat\textdagger & & Effect &  Sd. & Z-stat\textdagger \\
  \hline
  \textbf{Model 1}& $< $  1 yr & 136 & 29 & 4.68&& 172 & 11 & 16.34  \\
                           & $>$ 1 yr &  -11 & 43 & -0.38  && -12 & 16 & -1.13\\
                           & $\infty$   & $-\infty$  & -  & - &&$-\infty$ & - & -\\
   \textbf{Model 2}& $< $  1 yr & 320 & 31 & 10.44 && 322& 11 & 28.71  \\
                           & $>$ 1 yr & -15 & 46 & -0.49   &&-8 & 17 & -0.72   \\
                           & $\infty$   & $-\infty$  & -  & - &&$-\infty$ & - & -\\
    \textbf{Model 3}& $< $  1 yr & 327 & 31 & 10.64 && 325 & 11 & 28.81  \\
                           & $>$ 1 yr & -14 & 46 & -0.45   &&-7& 17 & -0.62  \\
                           & $\infty$   & $-\infty$  & -  & - &&$-\infty$ & - & -\\
     \hline
\textbf{Model 4}& $< $  1 yr & 362 & 17 & 21.60 && 378 & 6 & 61.35    \\
                           & $>$ 1 yr & 8 & 24 & 0.33   &&7 & 9 & 0.8   \\
                             & $\infty$   & $+\infty$  & -  & - &&$+\infty$ & - & -\\
\textbf{Model 5}& $< $  1 yr & 133 & - & -  && 136 & - & -     \\
                           & $>$ 1 yr & 84 & - & -   &&86 & - & -   \\
                            & $\infty$  & 359 & - & - && 370 & - & -\\
                            & Param. &149 & 5 & 31.24 && 154 & 2 & 89.36 \\
 \textbf{Model 6}& $< $  1 yr & 325 & - & -  && 334 & - & -    \\
                           & $>$ 1 yr &31 & - & -   &&34 & - & -   \\
                            & $\infty$ CD4   & 360 & - &-&& 371 & - & -\\
                           & $\infty$ VL   & -1.9 & - & - && -2 & - & -\\
                            & Param. CD4&600 & 21 & 28.42 && 630 & 8 & 82.22\\
                            & Param. VL&-7 & 0 & -40.86&& -7 & 0 & -120.51 \\
                                \hline
   \textbf{Model 7}& $< $  1 yr & 312 & - & -  && 304 & - & -    \\
                           & $>$ 1 yr &2 & - & -   &&4 & - & -   \\
                            & $\infty$ CD4   & 308 & - &-&& 306 & - & -\\
                           & $\infty$ VL   & -5.6 & - & - && -4.98 & - & -\\
        &Param. $\gamma$ & -1.12&            0.014&  -79.3 && -1.03&            0.003& -295.6  \\
\hline
    \multicolumn{9}{l}{ \textdagger Estimates for treatment effect ($\beta$) are significant at level 10\% if the absolute value of  Z-stat is} \\
       \multicolumn{9}{l}{ greater than 1.64 and significant at level 5\% if the absolute value of Z-stat is greater than 1.96.}\\
        \multicolumn{9}{l}{ $*$ To be compared with mean treatment effect in treated for ($<1$ year; $>1$ year; $\infty$): }\\
        \multicolumn{9}{l}{  benchmarks values are (350;12;370) for these simulations.}\\
   \hline
 \end{tabular}
\end{center}
\label{tab:resultsimul}
\end{table}

\section{Real data}
\label{part:real}
 We used two large cohorts: the ANRS CO3 Aquitaine cohort \citep{thiebaut2000clinical} and the Swiss HIV Cohort Study (SHCS) \citep{sterne2005long,gran2013}. Similarly to \cite{Cole2005}, we took a sub-sample of patients who were alive, HIV positive, yet untreated and under follow-up in April 1996 when HAART became available. All patients taking ARV in mono- or bi-therapy instead of HAART were excluded. Once a patient was on any therapy, we assumed he or she remained on it.  For each of them, the follow-up begins with the first visit after April 1996 and ends with 1) the last visit at which he or she was seen alive, 2) the last visit before patient discontinued the study, or 3) April 2003, whichever comes first. Data were supposed missing at random (MCAR); thus we deleted observations where either the viral load or the CD4 count was missing. Patients with at least 2 observations were included. See Web-Supplementary Material B1 for a more precise description. Finally, we considered 1591 patients the Aquitaine cohort and 1726 patients for the SHCS. Table \ref{tab:descriptionDataset} gives descriptive statistics.

\begin{table}[ht]
\centering
\caption{Data description for illustrations : Average viral load, CD4 counts and percentage of treatment attribution in the population are displayed for simulated data and real data from the Aquitaine cohort and the SHCS. Statistics displayed are mean [Q1;Q3].}
\label{tab:descriptionDataset}
\small
\begin{tabular}{lcccc}
  \hline
 & \multicolumn{2}{c}{Simulated}     & Aquitaine &   \\
 & \multicolumn{2}{c}{dataset}  & Cohort & SHCS\\
  \hline
  \textbf{n}&200&1500&1591&1726\\
  \hline
         \textbf{Missing data}  && & & \\
Administrative  & -&-& 81.6\%&  74.7\%\\
Death  & -&-& 12.7\%&  6.4\%\\
Lost of follow-up  & -&-& 5.7\%&  18.9\%\\
\hline
    \textbf{CD4 count}  && & & \\
 Baseline  &428  [ 266 ; 545 ]& 420  [ 253 ; 530 ]& 471  [ 298 ; 612 ] &536  [ 357 ; 670 ] \\
  Follow-up untreated &594  [ 485 ; 675 ] & 588  [ 478 ; 656 ]& 625  [ 440 ; 762 ] &543  [ 363 ; 675 ] \\
  Follow-up treated  & 627  [ 417 ; 837 ]& 606  [ 405 ; 801 ] & 492  [ 315 ; 638 ]& 507  [ 300 ; 660 ]\\
        \hline
  \textbf{Viral Load}  && & & \\
  Baseline & 3.9  [ 3.3 ; 4.6 ]&4  [ 3.4 ; 4.7 ]& 4.2  [ 3.6 ; 4.8 ] &4.0  [ 3.4 ; 4.6 ]\\
  Follow-up untreated &3.5  [ 2.9 ; 4.2 ] &3.7  [ 3.1 ; 4.4 ]& 3.3  [ 2.3 ; 4.2 ] & 3.8  [ 3.1 ; 4.5 ]\\
  Follow-up treated &2.6  [ 1.7 ; 3.2 ]& 2.6  [ 1.7 ; 3.2 ] & 2.7  [ 1.7 ; 3.6 ] &3.2  [ 2.4 ; 4.1 ] \\
  \% undetectable viral load&(3\%;4\%;40\%) & (2\%;3\%;38\%)& (7\%;22\%;48\%) & (10\%;15\%;57\%) \\
   (baseline,untreated, treated) &&&&\\
     \hline
    \textbf{Treatment attribution}  && & & \\
Time (day) & 412  [ 1 ; 631 ]&377  [ 91 ; 451 ]& 727  [ 1 ; 1281 ]& 548  [ 183 ; 752 ]\\
  \% treated & 69\% & 65\% & 64\% & 34\% \\
     \hline
\end{tabular}
\end{table}

 Table \ref{tab:resultCD4} displays the results we obtained for the treatment effect on CD4 counts.
The naive {Model 1}, not correcting for treatment attribution, indicated a small and non-significant increase of CD4 for SHCS cohort, and a significant negative effect for the Aquitaine Cohort; this illustrates the need for correcting for treatment attribution.
{Models 2 and 3} have different weights but show rather similar results, probably because treatment initiation was mainly driven by observed CD4 count.
Both models yielded a significant increase of CD4 counts for one year of treatment and after one year. The one-year increase, however, was much smaller for the Aquitaine cohort than for the SCHS. See Web-Supplementary Material B2 for a discussion of this result in relation with a possible practical violations of the experimental treatment assumption \citet{cole2008constructing}.
The results of the dynamical models, especially {Models 6 and 7}, were more consistent between the two cohorts.

\begin{table}[ht!]
\caption{ Estimated treatment effect on CD4 counts from real data of the Aquitaine cohort and SHCS: Model 1: Naive regression; Model 2: MSM with simple treatment model; Model 3: MSM with more complete treatment model; Model 4: simple dynamic model; 5: autoregressive model; Model 6: bivarariate dynamic model; Model 7: mechanistic model.}

\begin{center}
\begin{tabular}{llccccccc}
  \hline
   & &\multicolumn{7}{c}{Real Dataset observational studies} \\
  & & \multicolumn{3}{c}{SHCS}&$\quad$&\multicolumn{3}{c}{Aquitaine Cohort} \\\cline{3-5}\cline{7-9}
  Model & $\beta$ treatment&   Effect &  Sd. & Z-stat\textdagger & & Effect &  Sd. & Z-stat\textdagger \\
  \hline
  \textbf{Model 1}& $< $  1 yr & 6 & 16 & 0.34 && -94 & 12 & -7.55  \\
                           & $>$ 1 yr &  30 & 6 & 5.42   && 30 & 3 & 9.75\\
                           & $\infty$   & $+\infty$  & -  & - &&$+\infty$ & - & -\\
   \textbf{Model 2}& $< $  1 yr & 206 & 18 & 11.47 && 59 & 15 & 3.87 \\
                           & $>$ 1 yr & 54 & 8 & 6.67   &&41 & 5 & 8.03   \\
                           & $\infty$   & $+\infty$  & -  & - &&$+\infty$ & - & -\\
    \textbf{Model 3}& $< $  1 yr & 208 & 18 & 11.31 && 36 & 20 & 1.87  \\
                           & $>$ 1 yr & 50 & 9 & 5.79  &&53 & 5 & 9.62  \\
                           & $\infty$   & $+\infty$  & -  & - &&$+\infty$ & - & -\\
     \hline
\textbf{Model 4}& $< $  1 yr & 189 & 11 & 17.33 && 109 & 9 & 12.03    \\
                           & $>$ 1 yr & 73 & 16 & 4.54   &&55 & 13 & 4.28    \\
                             & $\infty$   & $+\infty$  & -  & - &&$+\infty$ & - & -\\
\textbf{Model 5}& $< $  1 yr & 26 & - & -  && 45 & - & -    \\
                           & $>$ 1 yr & 14 & - & -   &&19 & - & -  \\
                            & $\infty$  & 55 & - & - && 79 & - & -\\
                            & Param.  & 60 & 4 & 16.04 && 14 & 3 & 4.12\\
\textbf{Model 6}& $< $  1 yr & 73 & - & -  && 92 & - & -    \\
                           & $>$ 1 yr & 26& - & -  &&16 & - & -   \\
                            & $\infty$ CD4   &104& - & -&& 111 & - & -\\
                           & $\infty$ VL   & -2.0 & - & - && -2.3 & - & -\\
                            & Param. CD4 & 80 & 16 & 5 && 28 & 18 & 1.53\\
                          & Param. VL & -3.29 & 0.09 & -38.4 && -3.19 & 0.1 & -30.55\\
                                \hline
   \textbf{Model 7}& $< $  1 yr & 104 & - & -  && 71 & - & -    \\
                           & $>$ 1 yr &18 & - & -   &&9 & - & -   \\
                            & $\infty$ CD4   & 127 & - &-&& 86 & - & -\\
                           & $\infty$ VL   & -4.09 & - & - && -3.14 & - & -\\
        &Param. $\gamma$ &-1.73&            0.05&          -34.79& & -0.89&            0.01&          -85.77   \\
\hline
    \multicolumn{9}{l}{ \textdagger Estimates for treatment effect ($\beta$) are significant at level 10\% if the Z-stat is greater than 1.64} \\
       \multicolumn{9}{l}{  and significant at level 5\% if the Z-stat is greater than 1.96.}\\
    \hline
 \end{tabular}\end{center}
\label{tab:resultCD4}
\end{table}

 {Model 6} is interesting because it dissociates the effect of the treatment on CD4 count and on viral load. The estimated treatment effect on CD4 count was small in both cohorts and was non-significant effect for the Aquitaine cohort. In contrast, the effect on viral load was highly significant in both cohorts. This is consistent with the type of action of antiretroviral treatments: the increase of CD4 counts is essentially mediated by the decrease in viral load, which is the direct effect of antiretroviral treatments. Such biological knowledge is incorporated in {Model 7}, where the treatment acts on the infectivity parameter. In view of the Z-statistics obtained by a Wald test of the hypothesis $\beta=0$ in Equation (\ref{modeldose}), the power obtained in Model 7 seems to be very high (this was confirmed by a likelihood ratio test).
Moreover, Model 7 gives an insight into the value of the biological birth and death rates of cells during the infection (see Web-Supplementary Material B3 for details). The estimates from the two data sets are rather consistent in the sense that they have the same order of magnitude, although a formal comparison would show that several parameters are different.

\begin{figure}[h!]
  \centering
    \includegraphics[width=1.0\textwidth]{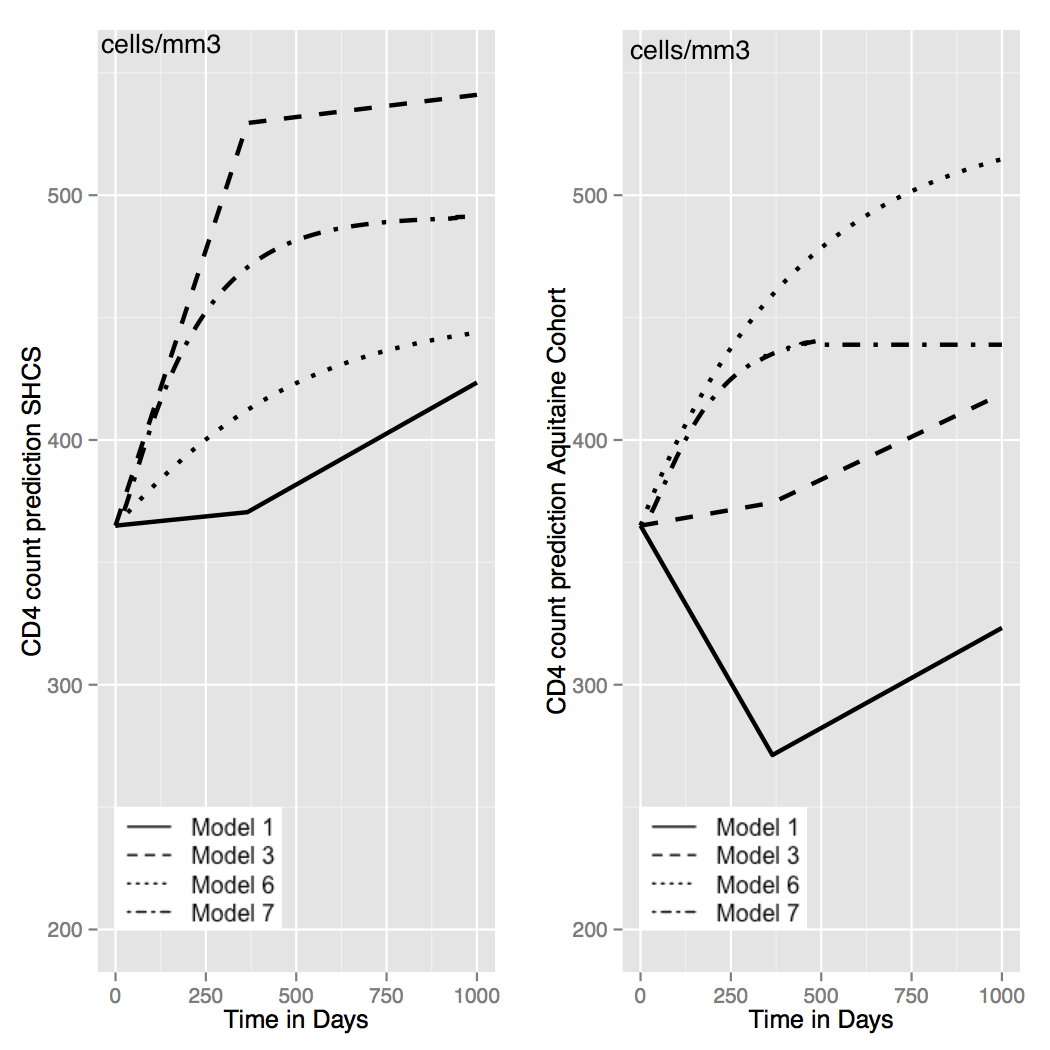}
      \caption{Mean evolution of CD4 predicted by Model 1 (plain line, simple regression), Model 3 (dashed line, MSM), Model 6 dotted line, linear incremental system) and Model 7 (dashed-dotted line, mechanistic model) for treated patients starting with 325 CD4 cells/mL and a viral load of 3.9 log10 copies/mm$^3$: (left) estimates from the SHCS data (right) estimates from the Aquitaine cohort.}
  \label{fig:meanCD4}
\end{figure}

Finally, a simple way to look at these results and to compare them, is to consider the mean evolution of CD4 along time. Figure \ref{fig:meanCD4} represents the predicted CD4 counts with Models 1, 3, 6 and 7 for treated patients starting at baseline with  CD4 count of 365 and a viral load of 4.4 (which are approximately the mean values at treatment initiation in the cohorts). For Models 1 to 3, these curves are deterministic, which is not the case for Models 4 to 7 that have random effects. For these latter models, we computed the mean predicted curves depending on the value of the random effect, which have to be set to values compatible with the baseline values of the biomarkers. In order to set them, in both case, we computed the equilibrium point of the system without treatment and solved the system of equations.
Figure \ref{fig:meanCD4} shows that the naive Model 1 badly under-estimated the effect. Both Models 1 and 3 are unstable, whereas Model 6 or the mechanistic Model 7 are more consistent between the two studies and have equilibrium points.

\vspace{-0.5cm}
\section{Conclusion}
\label{part:supp3}
This paper proposed four dynamic models for estimating the effect of HAART on CD4 counts and compared them to the naive regression model and two variants of previously proposed MSM models.
The naive regression model (Model 1) strongly underestimated the effect of the treatment. The MSM models (Models 2 and 3) corrected this misleading result but sometimes failed to reach significance or were unstable across studies. The discrete-time dynamic models (Models 4, 5 and 6) gave rather good estimates and appear to have a higher power, although they may sometimes be too rigid. All the discrete-time models can be easily fitted with classical softwares. The continuous-time dynamic model (Model 7) gave good results.
Models 6 and 7, which jointly model CD4 and viral load gave the most consistent results, with a richer interpretation since they take into account that the effect of HAART on CD4 is mediated by viral load.

 We have used a linear MSM with two slopes very similar to that proposed by \citet{Cole2005}. This model is adapted to represent the short term (few years) effect of treatment but not the long-term effect because it tends to infinity. It would be possible to define an MSM in which the effect would be bounded but this would be at the cost of additional non-linear parametrization. Also it would be possible to use more recent methods such as the history adjusted MSM \citep{petersen2007history} but their need is mostly justified to study dynamic treatment regimes whereas we assessed the effect of a static treatment regimes in this work. In contrast most dynamical models (although not Model 4) have an equilibrium point. Also, a MSM could be assumed for the increment $Z_t$ rather than for $Y_t$. Thus, the MSM approach could complement the dynamic approach in the sense that less stringent assumptions would be needed for causal inference. However, the dynamic models already do a good job and the need to correct them (at the price of more complex procedure and loss of power) is not obvious. In this paper we assumed MCAR observations for GEE which is justified by a majority of administrative censoring. MAR observations can be treated by using IPTC weights; see Web-Supplementary Material Section B4. The likelihood-based approach used for the dynamic models is valid for MAR observations.

 The mechanistic Model 7 directly incorporates biological knowledge. This leads to a more powerful test for the parameter of interest.  Moreover, it distinguishes the system living in continuous time and observations taken at discrete times. One of the advantages of this distinction is that we would be able to use all the data: with discrete-time models we must have approximately equally-spaced observations, which rarely occurs in real observational studies.  The simulated data in the paper are obtained with a dynamic model, so one might think that this favors dynamic models; however the true generation process comes itself from a complex system. Also, the simulated model is much more complex than Models 4-7 which are thus misspecified.


Finally, mechanistic models, once estimated, open the possibility of designing optimal (or sub-optimal) control of the therapy, as has been proposed on simulations by \citet{adams2004dynamic} and \citet{ernst2006clinical}, and also \citet{prague2012treatment}. The issue of ``optimal treatment regime'' has also been tackled outside of the context of mechanistic models  \citep{petersen2007individualized,orellana2010dynamic,saarela2015bayesian}. The drawback of the continuous-time approach is that it is numerically challenging and requires special software running on cluster computers.

\vspace{-0.5cm}
\section*{Web-Supplementary Materials}
Web-Supplementary Material referenced in Sections \ref{part:simul}, \ref{part:real} and \ref{part:supp3}, the simulated data analyzed in Section \ref{part:simul} and a R program implementing models 1-6 are available with this paper at the Biometrics website on Wiley Online Library. Programs to estimate the parameters with model 7 are available on a dedicated website: http://www.isped.u-bordeaux.fr/NIMROD.

\vspace{-0.5cm}
\section*{Acknowledgement}
The authors thank the investigators of the Aquitaine Cohort and the Swiss HIV Cohort Study. Parallel computing was used thanks to the MCIA (M\'esocentre de Calcul Intensif Aquitain) of the Universit\'e de Bordeaux and of the Universit\'e de Pau et des Pays de l'Adour.

\vspace{-0.5cm}
\bibliographystyle{biom.bst}
\bibliography{Causality_20150116}

\section*{APPENDIX: Correspondence between parameters of MSM and LIM}
The question is difficult for two reasons: the models are constructed differently; the philosophical approach to causality is different. We will make this exercise for comparing the MSM Models 2-3 and the dynamic Model 4. One can reconcile the two philosophical approach by saying that the ``causal'' interpretation (in a interventional point of view) is that, for a new patient who will be given treatment trajectory $\bar a_t$, we expect under the MSM models 2-3: $\Ee (Y_t(\bar a_t) | Y_{0})=  \beta_0 + \beta_1 \cum (\bar a_{t-1})+ \beta_2 \cumlag (\bar a_{t-1}) +\beta_3 t +\beta_4 Y_0$. Model 4 is formulated in terms of the increments $Z$: $Z_t=  \alpha_0 + \alpha_1 A_{t-1}+ \alpha_2 A_{t-2}+b+\varepsilon_t$, which by summation gives $Y_t=Y_0+ \alpha_1 \cum (\bar A_{t-1})+ \alpha_2 \cumlag (\bar A_{t-1}) +\alpha_3 t + M_t$,
where $M_t=b+\sum_{k=1}^t \varepsilon_t$ is a martingale. This is the Doob decomposition of the process $Y$.
With the assumption of a ``perfect'' system or a NUC system (see \citet{Commenges2015} and Chapter 9 of \citet{Commenges2015b}), if we apply treatment trajectory $\bar a_t$, this define a new probability $P^a$ under which the Doob decomposition is: $Y_t=Y_0+ \alpha_1 \cum (\bar a_{t-1})+ \alpha_2 \cumlag (\bar a_{t-1}) +\alpha_3 t + M_t$,
from which we deduce: $\Ee(Y_t|Y_0)=Y_0+ \alpha_1 \cum (\bar a_{t-1})+ \alpha_2 \cumlag (\bar a_{t-1}) +\alpha_3 t.$
Thus, Model 4 yields the same expectation under an intervention imposing treatment trajectory $\bar a$ as Models 2-3 if $\beta_4=1$, and the parameters giving the effects of $\cum (\bar a_{t-1})$ and $\cumlag (\bar a_{t-1})$ correspond. The way the model is estimated in the dynamic approach makes stronger assumptions than the MSM. Essentially it assumes that there is no confounder between $Z$ and $A$. That is, there is no variable that influences both $Z_t$ and $A_{t-1}$. For instance the viral load $V_{t-1}$ might be a confounder; such a confounder can be taken into account in the treatment model in a MSM. However, this confounding effect is not major; moreover, complex dynamic models (such as Model 6) can take viral load into account. For Models 5, 6 and 7, marginal effects can still be computed (analytically or by simulation), but this may lead to complex forms, while generally MSMs assume simple mathematical structures.

\label{lastpage}

\end{document}